\documentclass[sigconf]{acmart}
\AtBeginDocument{%
 }

\usepackage{hyperref}
\usepackage{fancybox}

\copyrightyear{2026}
\acmYear{2026}
\setcopyright{cc}
\setcctype{by}
\acmConference[CHI '26]{Proceedings of the 2026 CHI Conference on Human Factors in Computing Systems}{April 13--17, 2026}{Barcelona, Spain}
\acmBooktitle{Proceedings of the 2026 CHI Conference on Human Factors in Computing Systems (CHI '26), April 13--17, 2026, Barcelona, Spain}
\acmPrice{}
\acmDOI{10.1145/3772318.3790961}
\acmISBN{979-8-4007-2278-3/2026/04}

\begin{document}

\title[Participatory Threat Modeling with Chinese Young Women Living Alone]{``Create an environment that protects women, rather than selling anxiety!'': Participatory Threat Modeling with Chinese Young Women Living Alone}

\author{Shijing He}
\authornote{Both authors contributed equally to this research.}
\email{shijing.he@kcl.ac.uk}
\affiliation{
  \institution{King's College London}
  \city{London}
  \country{United Kingdom}
}

\author{Chenkai Ma}
\authornotemark[1]  
\email{chenkai.ma@kcl.ac.uk}
\affiliation{
  \institution{King's College London}
  \city{London}
  \country{United Kingdom}
}

\author{Chi Zhang}
\email{chi.zhang.9@warwick.ac.uk}
\affiliation{
  \institution{University of Warwick}
  \city{Coventry}
  \country{United Kingdom}
}

\author{Adam Jenkins}
\email{adam.jenkins@kcl.ac.uk}
\affiliation{
  \institution{King's College London}
  \city{London}
  \country{United Kingdom}
}

\author{Ruba Abu-Salma}
\email{ruba.abu-salma@kcl.ac.uk}
\affiliation{
  \institution{King's College London}
  \city{London}
  \country{United Kingdom}
}

\author{Jose Such}
\email{jose.such@csic.es}
\affiliation{
  \institution{INGENIO (CSIC-Universitat Politècnica de València)}
  \city{Valencia}
  \country{Spain}
}
\renewcommand{\shortauthors}{He et al.}

\begin{abstract}
As more young women in China live alone, they navigate entangled privacy, security, and safety (PSS) risks across smart homes, online platforms, and public infrastructures. Drawing on six participatory threat modeling (PTM) workshops (n = 33), we present a human-centered threat model that illustrates how digitally facilitated physical violence, digital harassment and scams, and pervasive surveillance by individuals, companies, and the state are interconnected and mutually reinforcing. We also document four mitigation strategies employed by participants: smart home device configurations, boundary management, sociocultural practices, and social media tactics--each of which can introduce new vulnerabilities and emotional burdens. Based on these insights, we developed a digital PSS guidebook for young women living alone (YWLA) in China. We further propose actionable design implications for smart home devices and social media platforms, along with policy and legal recommendations and directions for educational interventions.
\end{abstract}



\maketitle

\section{Introduction}
The rapid urbanization and increasing independence of women in China have led to a significant rise in the number of young women choosing to live alone, approximately 40 million, representing 2.83\% of the population~\cite{china2021population}. While this demographic shift reflects broader societal progress, it also amplifies existing gendered vulnerabilities, particularly given China's unique socio-legal environment, where institutional protections remain limited and digital visibility can heighten concerns regarding privacy, security, and safety (PSS)~\cite{he2025living}. Existing research often overlooks China's distinct sociotechnical context, where risks follow different cultural logics and coping strategies, and attending to these dynamics allows us to move beyond Western-centric assumptions toward more generalizable, context-sensitive understandings of PSS. In the Chinese context, these concerns are fundamentally shaped by traditional patriarchal values, where families are hetero-patriarchal and patrilineal, positioning female members in subordinate roles~\cite{Thornton1994}. The dualistic perspective of \textit{yin} (feminine) and \textit{yang} (masculine), originating from the \textit{Book of Changes} and reinforced by Confucianism~\cite{gao2003women}, has historically legitimized inequality between women and men. Beyond traditional values, the widespread adoption of digital surveillance technologies has introduced renewed PSS concerns and reinforced existing cultural stigmas surrounding women’s independence~\cite{he2025living}.

While Western feminist HCI often critiques smart homes as spaces where male co-inhabitants exert technological control~\cite{strengers2021smart,del2021controllable,despres2024my,chambers2020domesticating}, and studies from South-Asian contexts highlight kin oversight and shared-device use as key sources of abuse (e.g.,~\cite{sambasivan2019they,sambasivan2018privacy}, see \S\ref{threat_model_women}), a state-led framing of gender equality in China has expanded women's responsibilities without providing corresponding institutional support (see \S\ref{related_feminism_china}). As a result, vulnerability arises less from male dominance within the home and more from the absence of legal and community-based safety nets. Smart homes are thus perceived not only as convenience tools but also as safety infrastructures, meaning that so-called ``gender-neutral'' designs can unintentionally intensify gendered harms~\cite{he2025living}. We therefore situate this setting comparatively rather than treating it merely as a testing ground for existing work: while abuse categories and harm scales may exist across contexts, key actors, leakage points, and available forms of recourse differ. Recognizing this distinction helps move beyond Western-centric assumptions and better understand how PSS risks emerge when individual rights are structurally constrained. In this context, Chinese young women living alone (YWLA) are particularly vulnerable to gender-based violence, including violent crimes.\footnote{A man murdered a Chinese YWLA, and her body was transported in a suitcase in Shanghai 2021: \url{https://www.sohu.com/a/495679964_400421}.} For YWLA, harms stemming from the systemic absence of safety infrastructures are compounded by their departure from normative family structures and resulting social isolation, which further restricts access to legal, emotional, and communal support and heightens vulnerability to shame and fear~\cite{Zeng2020,rini2022deepfakes,laffier2023deepfakes,akter2025emergence,he2025living}.

While rising PSS concerns may contribute to the growing demand for personal protection tools, the adoption of smart home technologies in China has expanded rapidly, with over 78 million households using such devices~\cite{chinadata}. Existing literature has explored women's PSS concerns, including intimate partner surveillance, harassment in ride-hailing services, safety risks posed by smart home technologies, and the emergence of sexually explicit deepfakes~\cite{karusala2017women,ammari2022moderation,andalibi2018social,yoo2021anshimi,gallagher2019reclaiming,rho2018fostering,zou2021role,freed2017digital}. Yet few studies have examined how YWLA perceive PSS risks or what protective practices they adopt, particularly in non-Western contexts. Notably, He et al. demonstrated that smart home devices (e.g., smart locks, cameras, video doorbells) can simultaneously pose privacy and safety risks and serve as key mitigation strategies for Chinese YWLA~\cite{he2025living}. Building on these findings, we treat smart homes in this paper as theoretically motivated threat surfaces warranting focused and explicit discussion, while still allowing participants to surface additional threats and contexts.

To address this gap, we employed Participatory Threat Modeling (PTM)~\cite{slupska2021participatory}, a research approach that enables reinterpreting cybersecurity through a feminist lens, to explore YWLA's lived experiences, threat perceptions, and everyday mitigation strategies. PTM centers the lived experiences of individuals who may be regarded as marginalized and at-risk by inviting them to define their own PSS threats, reflect on their needs, and take action in communal workshop settings (see details in \S\ref{ptm}). To this end, we address the following research questions (RQs):

\begin{itemize}
\item \textbf{RQ1:} How do Chinese YWLA perceive PSS threats in their daily lives?
\item \textbf{RQ2:} What practices and mitigation strategies do these Chinese YWLA adopt to manage perceived PSS threats, and how do they view such practices and strategies?
\item \textbf{RQ3:} What improvements and recommendations are needed to ensure safe and equitable living environments for YWLA in China?
\end{itemize}

We designed and conducted six PTM workshops with Chinese YWLA ($n = 33$). Participants described a threat landscape defined by three key elements: (1) digitally facilitated physical threats posed by service workers (e.g., delivery drivers) who gained or could gain access to their homes; (2) digital harassment and scams, including deepfake extortion targeting them or their parents; and (3) pervasive surveillance, ranging from hidden cameras to state biometric systems. Participants viewed these threats as particularly severe because they were often initiated by men or male-dominated institutions and could rapidly escalate into offline, gender-based violence (see \S\ref{threat_model}). \textbf{(RQ1)}%

To navigate these threats, participants reported four main strategies: using smart home devices as ``digital guardians,'' performing masculinity and managing social and physical boundaries, drawing on traditional beliefs for psychological comfort, and relying on online communities for information and emotional support (see \S\ref{mitigation_practices}). However, participants also emphasized the ambivalence of these practices, in which smart homes introduced new privacy vulnerabilities, boundary work reinforced patriarchal norms and purity expectations, and social media coping strategies both supported and intensified anxiety through sensational, algorithmically amplified content. \textbf{(RQ2)}

Participants called for multi-layered interventions, including stronger legal and regulatory protections (e.g., higher penalties and clearer definitions for gendered digital and physical abuse) and more accountable data protection and privacy laws and regulations. They additionally stressed the need for community and educational reforms (e.g., professional training for law enforcement) and clearer, more usable privacy policies and platform mechanisms to make data use and processing, as well as safety tools, understandable for YWLA (see \S\ref{implications_suggestions}). \textbf{(RQ3)}

\textbf{Contributions.} We present a human-centered threat model using PTM with Chinese YWLA, showing how digitally facilitated physical harm, platform-mediated harassment and scams, and multi-scalar surveillance become mutually reinforcing under constrained control and uneven institutional support (see \S\ref{threat_model_framework}). By connecting threat pathways to feasibility constraints and responsibility allocation, we extend prior human-centered threat modeling work with an empirically grounded account of how gendered safety risks are produced and managed in a non-Western socio-technical setting. To translate insights into practical solutions, we present an accessible, open-access \textbf{Privacy, Security, and Safety (PSS) Guidebook} tailored specifically to young Chinese YWLA.\footnote{The PSS guidebook is freely accessible online at: \url{https://sites.google.com/view/pssguidebook}.} We further provide design recommendations for smart home companies (see \S\ref{suggestion_to_smart_home}) and social media platforms (see \S\ref{suggestion_social_media}), as well as actionable suggestions for policymakers (see \S\ref{policy_suggestions}).%

\section{Related Work}
\subsection{Emerging PSS Threats for Women}
\label{women_pss_concerns}

Research on women’s privacy, security, and safety (PSS) has highlighted the wide range of vulnerabilities they face across both public and private contexts. Public spaces often expose women to risks such as harassment, stalking, and assault~\cite{gardner2017harassment,rudman2021social,tandogan2016fear,karusala2017women}. At the same time, women encounter distinct PSS challenges in private settings, including sensitive data breaches, unwanted or undisclosed surveillance (e.g., domestic workers living in employer-controlled smart homes), and domestic violence (e.g.,~\cite{madriz2023nothing,freed2017digital,slupska2021threat,zou2021role,malki2024exploring,abu-salma2025grand}). These threats not only pose risks of physical harm but also have significant mental health consequences, contributing to increased anxiety, stress, and persistent feelings of vulnerability~\cite{hawkley2015perceived,hou2020gender,zou2021role}.

\subsubsection{Smart homes.}
Smart home devices have rapidly gained popularity, offering women increased convenience and a heightened sense of security~\cite{he2025living}. The ability to remotely monitor one’s home can feel empowering, enabling users to respond quickly to perceived threats and better manage everyday safety concerns~\cite{barbosa2019if}. At the same time, smart home technologies inherently involve trade-offs between privacy and security~\cite{thakkar2022would,he2025living,saqib2025bystander}. Users often tolerate a degree of surveillance in exchange for perceived security and safety benefits, reflecting a trade-off between the advantages of sharing personal data and the risks of privacy loss~\cite{kim2019willingness,lau2018alexa}. For women in particular, this trade-off is frequently viewed as acceptable when concerns about physical safety outweigh privacy considerations~\cite{strengers2019protection,slupska2021threat,del2021controllable}. Accordingly, prior work has proposed a range of technical interventions aimed at mitigating these risks (e.g.,~\cite{karusala2017women,paradkar2015all,sogi2018smarisa,yoo2021anshimi}).

At the same time, entrenched gender norms often position women as “passenger” users in smart homes, while men dominate the installation, configuration, and ongoing management of these technologies~\cite{strengers2019protection,koshy2021we,kraemer2020further,geeng2019s,del2021controllable,pink2023smart,chambers2020domesticating,despres2024my,he2025exploring}. Such dynamics can place women in vulnerable positions, as they may have limited control over systems ostensibly designed to protect them. This imbalance is further compounded by broader social perceptions that frame women as easier targets for crime, intensifying their exposure to harm despite the presence of security technologies~\cite{madriz2023nothing}.

However, most prior research has focused on Western contexts, often overlooking the cultural and social factors shaping privacy, security, and safety concerns in non-Western settings such as China, although a small but growing body of work has begun to address this gap. For example, He et al.~\cite{he2025living} examined the safety experiences of Chinese YWLA through analyzing RedNote posts, showing how smart cameras and doorbells are used alongside performances of masculinity to deter perceived threats. Their study also revealed how social media discourse can amplify fear and commodify anxiety, framing self-surveillance as empowerment while state and patriarchal systems fail to address the structural roots of gendered violence. However, this work leaves key privacy issues underexplored (e.g., vulnerabilities in smart homes and the role of deepfakes) and relies solely on second-hand data, without offering a human-centered threat model grounded in participants’ experiences.

\subsubsection{Deepfakes.}
Generative AI tools have rapidly shifted from niche technologies to widely accessible consumer systems capable of producing hyper-realistic synthetic media, including malicious content such as deepfakes~\cite{wagner2019word,wu2025understanding}. A significant portion of deepfake content is pornographic and overwhelmingly targets women~\cite{gieseke2020new,nadimpalli2022gbdf}, violating sexual autonomy by placing women in sexualized contexts to which they never consented~\cite{kugler2021deepfake,fido2022celebrity,akter2025emergence,maddocks2020deepfake}, reinforcing online harassment, and causing shame, isolation, and victim blaming~\cite{casagrande2025new}. Deepfakes are also used to justify stalking, intimate partner surveillance, and threats of violence~\cite{rini2022deepfakes}, and can result in financial harms such as blackmail and extortion. Moreover, the ubiquity of selfies and personal images makes young women particularly vulnerable to being inserted into pornographic deepfakes, contributing to significantly higher levels of fear compared to men~\cite{sippy2024behind}, while the persistence of such content online can severely disrupt safety, social standing, and economic stability~\cite{franks2018sex,akter2025emergence}. Long-term reputational damage may hinder employment opportunities and compel victims to engage in self-censorship or withdraw from digital spaces~\cite{laffier2023deepfakes,flynn2022deepfakes}.

\subsection{PSS Threat Modeling for Women} \label{threat_model_women}
Prior studies have emphasized the need for threat models grounded in real-world contexts and lived experiences, which clearly map pathways to harm and assign responsibility to relevant actors, such as platforms and vendors, rather than focusing solely on end-users. Usman and Zappala formalized four key components of such models: context, threats, protective strategies, and reflection. They argued that threat model elicitation should surface not only the harms people face but also the coping strategies they employ and the barriers they encounter, providing a holistic view~\cite{usman2025sok}. Furthermore, they highlighted that human-centered threat modeling differs from systems-oriented approaches by accounting for emotional responses, social limitations, and the need for broader societal change. Warford et al. extended this perspective by showing how the experiences of at-risk groups are shaped by structural inequalities and social norms, demonstrating why purely technical solutions are often insufficient~\cite{warford2022sok}.

In particular, prior empirical studies with women in non-Western contexts suggest that human-centered threat modeling should treat safety risks as the product of the interaction between local social norms, everyday technological infrastructures, and constraints on recourse, rather than as isolated technical attacks. Three recurring patterns have emerged from this work. First, the boundary between trusted and adversarial actors is often fluid, as risks can arise within intimate or household relationships, where practices such as credential sharing are not simply irresponsible but reflect trust norms or mobility constraints~\cite{alghamdi2015security}.

Second, women’s perceptions of threats and what counts as a feasible mitigation strategy are shaped by context-specific fears and expectations of propriety. Control often depends on defaults, transparency, and consent dialogs designed to allow privacy boundaries to align with social expectations, rather than assuming they can be individually chosen or strictly technically enforced~\cite{dev2020lessons}.

Third, harms are frequently amplified through gendered public communication channels, where participation itself is risky. Harassment, threats, and blackmail can become normalized, while low female participation and male reinforcement of abuse constrain what victims can safely disclose and how they can seek support~\cite{vashistha2019threats}.

In response, coping and mitigation practices often become situated and performative, especially under conditions of shared access and surveillance, where women balance social expectations of openness with a need for secrecy through strategic, audience-aware practices regarding shared devices~\cite{sambasivan2018privacy}. These dynamics also suggest that effective mitigation requires counter-abuse infrastructures that redistribute responsibility away from individuals and toward the broader ecosystem~\cite{sambasivan2019they}.

However, our study is grounded in the Chinese context and focuses on YWLA, who navigate a distinct “guardianship gap”—a dynamic underexplored in prior work. Rather than emphasizing shared phone use or credential-sharing practices, we develop a human-centered threat model situated within a multi-scalar socio-technical environment, in which smart home devices and digital platforms jointly shape digitally facilitated physical violence, scams, and surveillance.

\subsection{Gender Hierarchy, Social Media, and Feminism in China} \label{related_feminism_china}
Enduring patriarchal expectations continue to shape Chinese women’s decisions regarding marriage, childbirth, and employment, while also influencing public attitudes~\cite{gui2020leftover,ji2015between,gaetano2014leftover,yu2011varying,ge2014changes}. Recent state messaging promoting fertility reinforces these expectations by positioning women as bearers of the nation’s future~\cite{zhang2024one}. Within this context, women confronting sexual violence face persistent, culturally specific challenges~\cite{liu1999enduring,xu2005prevalence,wang2023some,Zeng2020}. Heightened monitoring of women’s behavior is often justified by narratives portraying them as vulnerable and in need of protection~\cite{cai2023ride}, frequently resulting in victim-blaming and a reliance on self-protection technologies, rather than addressing structural sources of gendered violence~\cite{sugiura2020victim,qin2024dismantling}.

Digital platforms play an ambivalent role in shaping how safety and feminism are articulated in China. On the one hand, they function as infrastructures for collective sensemaking, circulating safety-related news and enabling women, particularly younger cohorts, to share experiences, mobilize support, and demand change~\cite{mao2020feminist,huang2023anti,tian2024womb,Zeng2020}. On the other hand, these platforms also act as infrastructures of constraint. Prior work shows that heavy policing shapes which forms of safety discourse are visible, sustainable, and recognized as legitimate public discussion~\cite{xu2011internet,online_space_han_2018}. Consequently, feminist activists and KOLs often engage in careful boundary work, including self-censorship and strategic framing, to maintain community viability under regulation, while still facing misogynistic attacks and fear-laden narratives that can limit participation and undermine solidarity~\cite{peng2020feminist,xu2011internet}. Despite these constraints, online communities continue to support practical safety work, recent HCI research demonstrates how Chinese YWLA use online spaces not only to express fear and seek validation but also to co-develop creative self-protection strategies when institutional recourse is limited~\cite{he2025living}.

At the same time, feminism itself has become politically fragile, as it is often framed as a Western import, inviting suspicion and censure even when it emerges from local experiences of harm and inequality~\cite{huang2023anti}. Critics may portray feminist advocacy as social antagonism or as claiming “rights without responsibilities,” further restricting the types of safety demands that can be voiced without backlash and delegitimizing women’s accounts of risk~\cite{yang2023rural,mao2020feminist,deng2024persuasion,wu2019made,peng2020feminist}. In addition, commercial culture intertwines with feminist discourse, as brands frequently equate female empowerment and independence with consumption, blending political ideals with market logics~\cite{thornham2010just,wu2019made,yang2023post}. This fusion further complicates the landscape in which Chinese women navigate both their safety and their sociopolitical agency.

\subsection{Participatory Threat Modeling (PTM)} \label{ptm}
In this study, we employed PTM to elicit threat models, while explicitly acknowledging its priming effects and limitations. We connect protective strategies to the real-world constraints and legal, platform-based, and social challenges participants faced, and base our implications on participants’ own advice and visions, rather than relying solely on researcher-driven recommendations. PTM resists framing users as problems to be corrected and avoids narrowing solutions to purely technical fixes. Instead, it draws on participatory action research and security design~\cite{heath2018holding,yao2019defending,leitao2018digital,slupska2022they,slupska2021threat}, offering an open-ended, socially grounded method that emphasizes empowerment, collective resilience, and the socio-demographic realities shaping individual experiences and vulnerabilities.

Aligned with PTM principles, Wilkinson~\cite{wilkinson1999focus} and Slupska et al.~\cite{slupska2022they} argue that feminist research methods should prioritize participants’ lived experiences and affirm their agency. Slupska et al.~\cite{slupska2022they} note that researchers using PTM typically facilitate workshops where individuals, particularly those often overlooked in mainstream cybersecurity research, can reflect on what PSS means in their daily lives. Through collective modeling, participants identify perceived threats and articulate both existing and desired measures for protection, improved digital practices, and safer spaces.

\subsection{Contributions to Prior Work} \label{contributions}
Prior research has often overlooked the distinctive sociocultural, technological, and regulatory environments of China. Addressing this gap, our study makes three key contributions:

First, we present a human-centered threat model that highlights distinct actors and threat surfaces relevant to Chinese YWLA (digitally facilitated physical violence, digital harassment and scams, and pervasive multi-scalar surveillance) and illustrates how these threats intersect and reinforce one another. We further identify four clusters of double-edged mitigation strategies (smart homes, relationship and boundary management, sociocultural practices, and social media use) that can simultaneously reduce and redistribute risks.

Second, we adapted PTM to a non-Western, gendered context and connected it to Communication Privacy Management (CPM) theory. We used CPM to analyze how YWLA participants managed their privacy boundaries and coordinated with key stakeholders, highlighting issues of co-ownership and privacy turbulence (see \S\ref{rethinking_CPM}). In doing so, we build on He et al.'s work on security and safety among Chinese YWLA~\cite{he2025living} by moving beyond social media trace analysis to primary PTM workshop data collection and analysis, foregrounding PSS, and theorizing privacy boundary management beyond performative masculinity alone.

Lastly, moving beyond academic analysis of women's PSS concerns and mitigation strategies, we developed an open-access PSS guidebook in the form of a website based on insights from our PTM workshops (see \S\ref{pss_guidebook}). This practical resource combines low-tech preventive tactics, digital literacy guidance, and legal support information, directly responding to feminist HCI and PSS scholarship that calls for research outputs with real, tangible benefits for participants. In doing so, we shift from extractive research toward material reciprocity. Additionally, we offer multi-layered, actionable recommendations for a range of stakeholders, addressing both technical and social aspects.

\section{Methodology}
We conducted six PTM workshops with 33 Chinese YWLA, with each workshop involving 4–7 participants. All sessions were held via Microsoft Teams and lasted an average of 99.3 minutes. The translated workshop guidelines and questions, screening survey, follow-up survey for collecting feedback on the PSS guidebook, and the full codebook (all available in English) are open access and can be found at \href{https://osf.io/3dgpj/overview?view_only=97ed67f2208a4fd9a0e1c1c335bb8d7f}{this OSF repository}.%

\renewcommand{\arraystretch}{1}
\begin{table*}[!ht]
    \centering
    \fontsize{8pt}{10pt}\selectfont
    \resizebox{1\textwidth}{!}{%
    \begin{tabular}{llllllp{6cm}}
    \toprule 
        \textbf{ID} & \textbf{Age} & \textbf{Exp. living alone} & \textbf{Education} & \textbf{Occupation} & \textbf{Location} & \textbf{Owned smart home devices} 
        \\ 
    \midrule
        W1-01 & 30-35 & 3-5 years & Bachelor’s & Full-time & Shanghai & Smart TV, Security camera 
        \\
    \hline
        W1-02 & 23-29 & 3-5 years & Bachelor’s & Full-time & Chongqing & Smart lights, Smart speaker 
        \\
    \hline
        W1-03 & 30-35 & 1-2 years & Master's & Full-time & Beijing & Smart TV, Smart speaker, Smart doorlock/doorbell 
        \\
    \hline  
        W1-04 & 23-29 & 1-2 years & Master's & Student & Shanghai & Smart TV, Smart speaker, Smart doorlock/doorbell, Security camera 
        \\
    \hline
        W2-01 & 23-29 & 1-2 years & Master's & Full-time & Beijing & Smart speaker, Security camera 
        \\
    \hline
        W2-02 & 23-29 & 1-2 years & Bachelor's & Full-time & Zhejiang & Smart lights, Smart speaker, Security camera 
        \\
    \hline
        W2-03 & 23-29 & 3-5 years & Bachelor's & Full-time & Shanghai & Smart speaker, Security camera 
        \\
    \hline  
        W2-04 & 23-29 & 1-2 years & Bachelor's & Full-time & Anhui & Smart lights, Smart doorlock/doorbell 
        \\
    \hline  
        W2-05 & 18-22 & $<$ 1 year & Bachelor's & Student & Shandong & Smart speaker, Security camera 
        \\
    \hline  
        W2-06 & 23-29 & $<$ 1 year & Master's & Part-time & Guangdong & N/A 
        \\
    \hline
        W3-01 & 23-29 & 1-2 years & Assoc./Tech. & Self-employed & Beijing & Smart TV, Smart lights, Smart speaker, Smart doorlock/doorbell 
        \\
    \hline
        W3-02 & 23-29 & 1-2 years & Bachelor's & Full-time & Gansu & Smart TV, Smart doorlock/doorbell, Security camera, Smart kitchen appliances 
        \\
    \hline
        W3-03 & 23-29 & 3-5 years & Assoc./Tech. & Full-time & Sichuan & Smart lights, Security camera 
        \\
    \hline  
        W3-04 & 30-35 & 3-5 years & Bachelor's & Full-time & Fujian & Smart TV, Smart lights, Smart speaker, Smart doorlock/doorbell, Security camera 
        \\
    \hline  
        W3-05 & 23-29 & $>$ 5 years & Bachelor's & Self-employed & Liaoning & Smart TV, Smart speaker, Smart doorlock/doorbell, Security camera, Smart kitchen appliances 
        \\
    \hline  
        W3-06 & 23-29 & 1-2 years & Master's & Unemployed & Guangdong & Smart TV, Smart lights, Smart speaker, Smart doorlock/doorbell, Security camera 
        \\
    \hline
        W4-01 & 23-29 & 1-2 years & Assoc./Tech. & Self-employed & Gansu & Smart TV, Smart lights, Security camera, Smart kitchen appliances 
        \\
    \hline  
        W4-02 & 30-35 & $>$ 5 years & Master's & Full-time & Inner Mongolia & Smart TV, Smart lights, Smart speaker, Smart doorlock/doorbell, Security camera 
        \\
    \hline  
        W4-03 & 23-29 & $<$ 1 year & Bachelor's & Student & Fujian & N/A 
        \\
    \hline  
        W4-04 & 30-35 & $>$ 5 years & Master's & Full-time & Shanghai & Smart lights, Smart speaker, Smart plug, Smart doorlock/doorbell, Security camera 
        \\
    \hline  
        W5-01 & 23-29 & 3-5 years & Bachelor's & Full-time & Guangdong & Smart lights, Security camera 
        \\
    \hline  
        W5-02 & 23-29 & 1-2 years & Assoc./Tech. & Full-time & Shandong & Smart lights, Smart speaker, Smart doorlock/doorbell, Security camera 
        \\
    \hline  
        W5-03 & 23-29 & 1-2 years & Master's & Full-time & Shanghai & Security camera 
        \\
    \hline  
        W5-04 & 30-35 & $>$ 5 years & Master's & Full-time & Jiangxi & Smart TV, Smart lights, Smart speaker, Security camera 
        \\
    \hline  
        W5-05 & 23-29 & 1-2 years & Bachelor's & Full-time & Zhejiang & Smart speaker, Smart doorlock/doorbell, Security camera 
        \\
    \hline  
        W5-06 & 30-35 & $<$ 1 year & Bachelor's & Self-employed & Guangdong & Smart lights, Smart doorlock/doorbell, Security camera, Smart smoke monitor/alarm 
        \\
    \hline  
        W5-07 & 30-35 & 1-2 years & Master's & Full-time & Guangdong & Smart TV, Smart doorlock/doorbell, Security camera 
        \\
    \hline  
        W6-01 & 23-29 & 1-2 years & Assoc./Tech. & Full-time & Zhejiang & Smart lights, Smart speaker, Smart doorlock/doorbell, Security camera 
        \\
    \hline  
        W6-02 & 23-29 & $>$ 5 years & Master's & Full-time & Henan & Smart TV, Smart lights, Security camera 
        \\
    \hline  
        W6-03 & 18-22 & $<$ 1 year & Assoc./Tech. & Part-time & Anhui & Smart TV, Smart speaker, Smart doorlock/doorbell, Security camera 
        \\
    \hline  
        W6-04 & 30-35 & 3-5 years & Assoc./Tech. & Full-time & Hunan & Smart TV, Smart lights, Smart speaker, Security camera 
        \\
    \hline 
        W6-05 & 23-29 & 3-5 years & Bachelor's & Full-time & Shandong & Smart lights, Smart speaker, Security camera, Smart kitchen appliances 
        \\
    \hline  
        W6-06 & 30-35 & $>$ 5 years & Master's & Full-time & Shanghai & Security camera 
        \\
    \bottomrule 
    \end{tabular}}
    \caption{Participant background information and demographics (Chinese YWLA), including experience living alone, education level, occupation, work and residence locations, and ownership of smart home devices (n = 33).}\label{tab:demographic}
\end{table*}

\subsection{Participant Recruitment}
We aimed to recruit a diverse participant sample in terms of demographics, backgrounds, and perspectives. Recruitment advertisements were distributed on social media platforms, including WeChat, Baidu Tieba, Douban, and RedNote. We also employed a snowball sampling strategy~\cite{parker2019snowball}. All potential participants completed a screening survey prior to inclusion.

We did not require prior experience with smart home devices or AI-based systems (e.g., GenAI), or AI-generated content (e.g., deepfakes), which helped avoid sampling only highly tech-experienced individuals and enabled us to capture variation among novice and early adopters~\cite{usman2025sok}. To be eligible, participants were required to: 1) be located in mainland China; 2) have current or prior experience living alone in mainland China; and 3) be aged 18 to 35.\footnote{The National Medium and Long-term Youth Development Plan (2016–2025) defines youth as individuals aged 14 to 35 in mainland China~\cite{youth_definition}. For ethical reasons, we defined the youth cohort in this study as individuals aged 18–35.}

Each participant received a 200 RMB gift card (\$27.6) as compensation for their time. Table~\ref{tab:demographic} describes participants’ background information and demographics.

\subsection{PTM Workshops}
Each workshop began with open-ended elicitation before introducing any technology-related prompts. Only after this initial phase did the moderator optionally present neutral, situational prompts to help minimize interpretive bias. We distinguished between threats that emerged organically during the open-ended elicitation (primarily concerning physical safety and surveillance) and those that arose after prompts related to smart home technologies and GenAI (including privacy violations and deepfakes).

Following a brief introduction in which the moderator outlined the study aims and ground rules, participants engaged in a short warm-up activity. During this activity, they shared their initial thoughts on what ``sense of safety'' meant to them while at home alone and described how they personally defined privacy and security.%

The main discussion was organized around three topics in accordance with our workshop guide: 1) perceived benefits and trust in smart home devices and other technologies, including motivations for adoption or avoidance, data privacy concerns, and reactions to potential security incidents such as data leaks; 2) perceived vulnerabilities and harms, identifying typical scenarios and psychological impacts related to feeling unsafe or insecure, exploring experiences with online advice, and evaluating awareness of and concerns about abusive online content such as deepfakes; and 3) protective mitigation strategies and desired interventions, examining how specific incidents had influenced participants' behaviors and their preferred protective measures, including discussions of responsibility for privacy at individual, enterprise, and governmental levels.

Participants then shared recommendations and prioritized the single most impactful change to enhance their sense of safety from technical, social, and legal perspectives. The workshop concluded with a brief wrap-up segment, allowing participants to offer any additional insights or comments.

In addition, we developed a dedicated PSS guidebook for Chinese YWLA (see \S\ref{pss_guidebook}) and distributed it to the same participants through a follow-up questionnaire to gather feedback for further iteration and improvement. Based on their input—such as revising the digital harassment and scam scenarios to better reflect the nuanced risks Chinese YWLA might face—the three authors refined the initial version through multiple rounds of collaborative discussions, focusing on clarity, practical applicability, and alignment with the lived experiences shared during the workshops. The finalized guidebook was subsequently published online for free access.

\subsubsection{Pilot Workshop}
Before conducting the main PTM workshops, we held a pilot workshop with four Chinese YWLA to ensure the questions were clear and to identify any potential issues in the guide early on. Data from this pilot workshop was not included in the final analysis.


\subsection{Research Ethics}
We took deliberate steps to ensure our study met ethical standards, including careful consideration of how to categorize Chinese YWLA. While some might view them as vulnerable or marginalized, we did not classify them as an at-risk population—unlike groups facing state-sanctioned harms or systemic discrimination (e.g., undocumented migrants~\cite{guberek2018keeping})—and this framing was affirmed by participants themselves, many of whom rejected the label of vulnerability. Nevertheless, we recognized their unique sociocultural challenges and treated them with the same care and protections recommended for vulnerable or marginalized groups~\cite{bellini2024sok}. Ethical issues were regularly discussed by the research team to ensure participants felt respected, safe, and fully supported. We also prioritized participant anonymity and obtained informed consent.


Only researchers involved in data collection had access to personal information. Before joining the study, each participant received a detailed information sheet, provided written consent, and had the opportunity to ask questions. To mitigate state surveillance risks, we avoided collecting unnecessary identifiers, allowed participants to join without video if they preferred, and stored all audio recordings and consent materials on encrypted, password-protected servers in the UK. During the workshops, participants used assigned IDs (e.g., W1-01) to maintain anonymity and were reminded that they could skip questions or take breaks at any time. Sessions were moderated by the third author (a woman) to enhance comfort and safety, with the first and second authors assisting with note-taking and timekeeping.

This study was approved by the Research Office at King’s College London (ID: MRSP-24/25-47331).

\subsection{Data Analysis}
We audio recorded and transcribed all workshop sessions. Author 1 then reviewed each transcript, corrected inaccuracies, and removed any identifiers (e.g., names). Authors 2 and 3 independently verified these revised transcripts to minimize any remaining errors. Notably, Authors 1–3 are all native Mandarin speakers born and raised in Mandarin-speaking regions and served as the primary data analysts. They are trained researchers with expertise in PSS, gender HCI, and qualitative data analysis, ensuring nuanced interpretation of cultural and technical terms. All transcripts were coded and analyzed in Mandarin to prevent potential loss of meaning during translation; Author 1 then translated the codes and quotes from Mandarin into English, which the other authors reviewed for accuracy.%

We reached data saturation—commonly defined as the point at which no new information or insights emerge—after the fifth PTM workshop. To confirm this, we conducted an additional workshop, which yielded minimal novel insights. While data saturation was observed, we acknowledge that workshop dynamics might influence the breadth of ideas and that alternative participant compositions could have surfaced different perspectives. Moreover, the concept of data saturation remains debated and inherently subjective~\cite{braun2006using,braun2021saturate}.

To address this, we adopted a reflexive thematic analysis approach~\cite{braun2006using}, continuously and inductively reviewing field notes and insights throughout data collection. Three authors performed iterative open coding, developing and refining codes. Transcript excerpts and field notes were tagged with codes derived from patterns in the data rather than a predefined scheme. The team then reviewed and organized the collective codes into themes and subthemes to address the study’s RQs, carefully examining participant excerpts to define and refine how each theme was represented. Themes were iteratively developed, refined, and collaboratively validated until the research team agreed that sufficient depth had been achieved to address the RQs~\cite{guest2006many}.%

\subsection{Limitations}
We relied on self-reported data from workshops, which might have introduced social desirability bias~\cite{nederhof1985methods}, as participants might underreport stigmatized experiences (e.g., harassment) due to cultural norms related to privacy and shame~\cite{owens2006conversational}. We employed PTM to explore in depth the PSS lived experiences of participants~\cite{slupska2022they}, focusing on neutral, open-ended prompts to help reduce bias. However, our subsequent emphasis on technology-specific prompts, based on prior evidence about YWLA~\cite{he2025living}, might have limited the spontaneous emergence of other topics—though some additional issues, such as physical violence facilitated by digital platforms (\S\ref{threat_model}), did emerge beyond the prompts.

Our cross-sectional design captured a snapshot of perceptions and practices but could not assess longitudinal changes in risk exposure or coping strategies~\cite{spector2019not}. Given the rapid development of smart home devices and AI-based systems, future work could prioritize longitudinal research, such as ethnographic studies tracking participants’ evolving risk perceptions and mitigation strategies over time (e.g.,~\cite{burrows2018privacy}) to uncover dynamic interactions between technology adoption and gendered vulnerabilities.

While we developed an open-access PSS guidebook, future research could examine its long-term effectiveness in addressing the PSS concerns of Chinese YWLA. We promoted our findings and the guidebook on social media platforms (e.g., RedNote). The guide draws on Chinese cultural repertoires and is designed for a Chinese audience; its applicability in other contexts warrants further exploration. This could include translating the guide into multiple languages and conducting cross-cultural studies to better understand and compare PSS threat perceptions and coping strategies across populations and sociocultural contexts (\S\ref{threat_model}), particularly where norms around privacy and surveillance differ.

Finally, while our findings shed light on the intersections of the threats of smart home ecosystems and deepfakes, we did not empirically trace how device-specific data leaks enabled synthetic harmful or sexual content, leaving causal links to be explored by future technical forensic analyses.

\section{Results} \label{result}

\subsection{Gendered and Digitally Mediated Threats}\label{threat_model}

\subsubsection{Actors and Attacks: Physical Violence Facilitated by Digital Platforms}\label{gender_threats}
Unlike generic threat models for women, the threat landscape for YWLA is shaped not only by their living arrangements but also by safety infrastructures and social norms that implicitly assume the presence of co-residents or ``protectors''—often male partners or family members—as informal buffers at household boundaries. Many of the threats our participants described (e.g., risks posed by delivery workers or untrusted co-inhabitants) were also experienced by women who did not live alone; however, participants emphasized that prevailing assumptions of guardianship altered both the likelihood and consequences of such threats. In the absence of socially expected co-residents, participants anticipated navigating these encounters on their own. This perceived ``guardianship gap'' thus reflected not a natural absence of protection, but a system that shifted responsibility for security onto women who lived independently, intensifying the burden of continuous vigilance and boundary management. As a result, participants expressed significant concern about physical threats posed by men, including strangers (e.g., passersby), acquaintances (e.g., landlords or roommates), and service providers (e.g., delivery workers).

\textbf{Male Service Providers and Address Leakage.} China's delivery workforce is predominantly composed of young male migrant workers~\cite{sun2019your}. These workers often gain access to users' home addresses and phone numbers through service platforms. Because a single courier typically serves a specific neighborhood, many gig workers become familiar with residents' contact details and locations. Consequently, a key concern for our participants was the unavoidable interaction with male gig workers (e.g., food or parcel couriers).

Although the interaction was digital, the threat posed was primarily physical. Some participants described disputes with delivery workers as particularly frightening due to living alone, as digital platforms eliminated the anonymity that might otherwise protect them in public. For instance, W1-02 recalled a conflict with a courier and feared retaliation: \textit{``He knew my address, and I lived alone, so I feared he might come back.''} Similarly, W1-03 expressed concern about maintenance workers identifying solo women as targets: \textit{``Some male maintenance workers might notice you have money and know you're living alone. If someone has bad intentions, it could lead to dangerous outcomes.''}

\textbf{Male Co-Inhabitants and Contextual Threats.} Participants shared experiences illustrating how shared spaces, especially with mixed-gender occupancy or unclear security arrangements, often induced heightened anxiety and feelings of vulnerability. This was frequently because many of our participants (like YWLA in China) rented rooms in shared housing due to financial constraints~\cite{harten2021housing}. Insecurity was further amplified by male co-inhabitants in these settings, where blurred domestic boundaries and gendered spatial norms generated persistent discomfort. W2-03 described a male roommate who would walk through shared spaces nearly naked after showering, violating normative expectations of bodily privacy. She also raised concerns about personal safety, stating: \textit{``I could hear him come in late at night, often drunk and loud, which made me feel unsafe because only a flimsy wooden door separated us.''} This highlights a failure of the home as a sanctuary: the threat actor was effectively inside the perimeter, bypassing the external security measures participants typically relied on.

\textbf{Strangers and Offline Signaling Attacks.} Several participants reported unexplained marks or signals appearing at their doors, which prompted intense concern about potential targeting by criminals, often interpreted through knowledge from online safety communities. As W3-02 shared: \textit{``I once noticed a strange X mark on my door. I really worried because I'd seen posts online saying some rapists mark doors to identify targets.''} This demonstrates how participants continuously scanned their physical environment for signs of targeted violence, a cognitive burden intensified by their isolation.

\subsubsection{Harms: Digital Harassment and Scams}\label{scam_threats}

While concerns about harassment emerged organically, discussions regarding GenAI (e.g., deepfakes) were primarily elicited through targeted workshop prompts. Participants’ responses to these prompts, however, revealed a distinct prioritization of harms: rather than focusing solely on the reputational damage of deepfake pornography~\cite{korea2024deepfake,mcglynn2025new}, they emphasized relational and financial harms (e.g., frauds and scams) targeting their parents.

\textbf{Deepfake Scams and Filial Piety.} Participants expressed acute anxiety that their status as YWLA living away from home created an informational void that scammers could exploit. They linked concerns about deepfake scams to a sense of responsibility for protecting their parents or elders, highlighting how digital vigilance was shaped by culturally embedded norms of filial piety~\cite{deng2025auntie}. Because parents could not physically verify the safety of their daughters (our participants), they were uniquely vulnerable to AI scams or GenAI fraud (e.g., voice cloning) claiming that their daughters were in danger. W1-03 noted that she was primarily concerned about deepfake scams because they could harm her parents rather than herself: \textit{``I always remind my parents about AI deepfake fraud. Even if it looks like my face and sounds like my voice, you can't assume the video is really me.''}

Some participants employed protective strategies against deepfake scams, particularly to safeguard family members vulnerable to deception by impostors. W2-03 emphasized: \textit{``Anyone claiming to be me requesting money must first verify their identity using secret questions only my parents and I know. Like how old I was when I fell off a wall and landed in the ER--very specific personal details.''} This underscores how participants had to perform additional emotional and logistical labor to protect their family members from threats posed by their own digital likenesses.

\textbf{Sexual Harassment and the Burden of Proof.} Regarding digital sexual harassment, participants highlighted how digital platforms blurred the boundaries of consent. W2-03 recounted a delivery driver adding her on WeChat after a purchase, leveraging the platform's data handoff to attempt an unwanted hookup. Similarly, W6-03 reported receiving explicit images via AirDrop on the subway. When prompted about deepfake pornography, participants expressed resignation regarding the legal system, noting that harm was not only inflicted by deepfakes but also by the institutional burden of proof. W1-02 further addressed the problematic practice of victim blaming in cases related to deepfake pornography, highlighting how current legal frameworks often placed an undue burden on victims, thereby exacerbating the harm inflicted by digital harassment and abuse: \textit{``The video itself is already a violation. Law enforcement should take it seriously without making me prove my innocence. Forcing women to do so only deepens the insult.''}

\subsubsection{Multi-Scalar and Gendered Digital Surveillance}\label{surveillance_threats}
Participants identified digital surveillance as a significant threat, describing it as both multi‑scalar and gendered. They distinguished between individual, corporate, and state forms of surveillance, revealing a broader conceptualization beyond traditional state‑centric frameworks. However, participants simultaneously feared private surveillance while tolerating state surveillance as a necessary substitute for personal guardianship.

\textbf{Individual-Level Digital Surveillance.} Surveillance at the individual level emerged as the most acute concern, with particular emphasis on micro-hidden cameras covertly installed by males in public spaces such as restrooms and hotel rooms. For example, W2-03 stated: \textit{``I started feeling anxious after hearing about hidden cameras in shared bathrooms. The friendlier my male roommate seemed, the more uneasy I became. Before each shower, I found myself spending several minutes checking every shelf.''} Similarly, W3-02 expressed significant concern about hidden cameras potentially installed in hotel rooms, which led to persistent anxiety and heightened self-surveillance behaviors. Additionally, W3-05 described a troubling incident in which her facial data was harvested without informed consent by a male acquaintance under the guise of a playful app purportedly designed to \textit{``match your face to a celebrity.''} The individual subsequently used this illicitly obtained data to stalk and locate her residence, highlighting the profound risks associated with the misuse of facial recognition technology.%

\textbf{Enterprise-Level Digital Surveillance.} Participants articulated a complex relationship with surveillance technologies deployed by enterprises. Central to their anxieties were the opaque and non-consensual nature of data collection practices and the potential misuse of personal data by manufacturers or engineers without users' awareness or permission. As W2-04 described: \textit{``Engineers can even communicate with you through the camera, or you might suddenly hear a stranger’s voice coming from it at home.''} W3-02 reinforced these concerns by emphasizing the troubling lack of transparency regarding the use and protection of data collected by smart home devices.

Some participants further highlighted worries about the increasing sophistication and commercialization of covert surveillance technologies, such as micro-hidden cameras. They criticized manufacturers for prioritizing profit and ease of use over the privacy and security of vulnerable individuals. W2-03 illustrated these anxieties: \textit{``Some companies go to great lengths to make them tiny and easy to purchase. This effectively makes them accomplices to the male criminals who took such photos of women.''} Participants’ accounts resonate with Zuboff's critique of surveillance capitalism, in which companies extract behavioral data under the guise of service provision~\cite{Zuboff2019}. This commodification of personal data exploits gendered fears and insecurities while offering users little transparency or meaningful recourse.

\textbf{State-Level Digital Surveillance.} Participants generally perceived state-level surveillance as an essential safety measure, particularly on streets and subways, as well as in gated compounds when walking home alone at night. Many deliberately chose \textit{``routes with more cameras''} to feel a greater sense of security. However, some noted that public CCTV coverage neither prevented incidents nor triggered visible interventions. For example, W6-02 observed: \textit{``These CCTVs only provide evidence after an incident occurs; they don’t actually protect you in real time.''}

This created a paradoxical tension in which surveillance simultaneously provided reassurance and provoked anxiety due to constant visibility and underlying power dynamics rooted in patriarchal governance structures. W3-06 elaborated: \textit{``I used to see CCTV cameras as a form of protection, but now it feels more like constant surveillance--someone always watching, limiting your actions and freedom. And most government officials are men.''} Additionally, W5-07 connected her concerns about personal security and surveillance to China’s COVID-19 pandemic data collection practices~\cite{liu2021privacy}, arguing that the grid-management system blurred the boundaries between public health measures and permanent surveillance infrastructure~\cite{mittelstaedt2022grid}. She explained: \textit{``The mass collection of personal data makes me panic--you never know how it might be used. Even after the pandemic, biometric checks are everywhere, with cameras requiring facial scans for basic services like subway entrances.''}

\noindent\shadowbox{\parbox{\linewidth}{\textbf{Key Findings (RQ1):} Participants identified three key threat categories grounded in their lived experiences: 1) digitally facilitated physical threats, such as delivery and rental platforms exposing YWLA to male service workers, landlords, roommates, and strangers who could access or closely monitor their living spaces; 2) digital harassment and scams, including deepfake scams, which significantly exacerbated mental stress and posed potential physical risks; and 3) digital surveillance, including unauthorized practices—both institutional (e.g., state biometric data collection) and personal (e.g., hidden cameras)—that heightened participants' anxiety about personal data misuse.

Notably, digital threats were perceived as more severe due to their potential to escalate into real-world physical harms, particularly because these threats were often initiated by men or by authorities operating within patriarchal norms.}}

\subsection{Mitigation Strategies and Best Practices} \label{mitigation_practices}
\subsubsection{Technological Substitution: Smart Homes as Digital Guardians} \label{mitigate_smart_home}

Most participants viewed smart homes positively, emphasizing their role in enhancing personal safety. They described smart home devices not only as tools of convenience but also as technological surrogates for physical guardians they lacked (e.g., family members or roommates). Some participants highlighted the proactive use of security cameras to deter theft--such as monitoring parcels left at the door--and underscored the dual convenience and security benefits of smart door locks. These locks enabled them to identify visitors before opening the door and to capture images of suspicious individuals that could later be provided to the police.

Despite acknowledging the utility of smart home devices, several participants expressed reservations about potential PSS risks inherent in these technologies. For example, W4-03 raised concerns about privacy violations stemming from opaque data-handling practices: \textit{``Surveillance cameras can record footage without consent and may even sell this data to third parties, including entities on the dark web.''} To mitigate such risks, some participants reported adopting strategies such as preferring local over cloud storage and physically disconnecting devices when privacy was paramount.

In addition, some participants demonstrated a strong awareness of bystander privacy, emphasizing the rights of individuals inadvertently captured by others’ smart home surveillance devices. They highlighted the responsibility of device owners to manage and limit data collection practices. For instance, W1-02 described her discomfort when passing through a neighbor’s monitored area: \textit{``Whenever I pass by, I instinctively turn my head away, convinced that it’s filming me.''} Similarly, W2-05 reflected on the ethical complexity of surveillance devices installed for pet monitoring that nonetheless captured detailed information about others’ routines when a feeder and camera were temporarily placed in a friend’s apartment: \textit{``I could see exactly when my friend walked by, what they were wearing, and even hear the room--it felt like spying on someone in \emph{The Truman Show}.''}

These tensions were particularly acute for participants renting single rooms within shared apartments, where control over public spaces and device permissions was often shared or uneven. Several participants described smart door locks whose primary accounts were held by landlords, roommates, or even digital rental platforms (e.g., Ziroom\footnote{Ziroom (also known as Ziru) is a major Chinese O2O platform offering long-term apartment rentals and integrated property management services across multiple cities.} [W4-04]), granting them unilateral access to entry logs or the ability to generate new passwords. As W3-04 explained, \textit{``The smart door lock is right outside my flat, and I can use it, but it’s also linked to my landlord’s app [...] He can see exactly when I leave or get home, and he can even let himself in with a temporary password.''} Similarly, W5-03 described how her parents purchased and remotely configured a camera outside her doorway with protective intentions: \textit{``It reassures them, but sometimes it makes me feel like I’m living under their watch again.''}%

\subsubsection{Relationships and Boundary Management} \label{mitigate_relationship}

Many participants described a deliberate and conscious effort to navigate relationships with men carefully, often adjusting their interactions to avoid conflict and minimize exposure to risk. For example, W1-03 reported adopting a preventive strategy with Didi drivers--consistently sitting in the back seat--after experiencing an incident of sexual harassment that left her feeling disrespected. Several participants further noted that their strategies for managing relationships and concealing aspects of their identity were influenced by the happiness principle~\cite{happiness_principle}, which they linked to the traditional Chinese proverb: \textit{``Endure for a moment, and the storm will pass; step back, and the sea and sky will stretch wide before you.''}

Nevertheless, W5-07 critically reflected on how such concessions could reinforce the gender hierarchies they were intended to circumvent, while acknowledging the compromises necessitated by the difficulty of reforming patriarchal norms in China: \textit{``Following it suppresses my needs, reinforces stereotypes, and undermines gender equality. Structural reform may be costly, but I believe this principle can reduce my risk at little expense by softening feminine cues, so I accept the trade-off.''}

Many participants described deliberately simulating a male presence--for example, using clothing and footwear as props--to suggest male occupancy and deter unwanted attention or harassment. However, W2-05 reflected on these tactics: \textit{``fake security actions are mostly about self-comfort.''} She emphasized that such practices often functioned as psychological coping mechanisms rather than robust safeguards, providing confidence but insufficient protection against serious threats.


Many participants frequently used ``Mr.'' in their online profiles, adopted typical Chinese male names or avatars on social media, or used male voices (e.g., using their male family members' voices to simulate a bustling household) to obscure their gender identity and enhance their safety. However, some participants extended this practice to linguistic style-shifting. For instance, W4-03 leveraged a stereotype to discourage aggressive bargaining, stating: \textit{``While I'm pretending to be a man--I don't have to follow the usual niceties--punctuation, polite phrasing, that sort of thing. I can type chaotically, make random mistakes--and it feels manly.''} Interestingly, W5-03 noted that her gender disguise extended beyond the digital context, permeating deliberate personal presentation and public behavior, stating: \textit{``Although I admit that women are more vulnerable than men, these behaviors do not affect my gender identity [...]. If I go out alone, I avoid revealing clothes. I'm fairly tall, and in loose outfits you can't tell from behind whether I'm male or female. This simple trick lowers my risk, and that's enough.''} These tactics reflected participants' strategies to behave as men as a pragmatic means to deter harassment and signal male presence. This constitutes \textit{performing under duress}: participants' disclosed gender identity was shaped not by self-expression, but by fear and the imperative of self-protection~\cite{butler2015notes}.

\subsubsection{Chinese Social Practices and Cultural Beliefs} \label{mitigate_cultural}
We found that Chinese social norms and cultural beliefs significantly influenced participants’ perceptions of, and strategies for mitigating, PSS. Traditional Chinese ideals of female virtue and purity (Qingbai) cast any sexual contact--consensual or forced--as a stain on a woman’s moral worth and her family’s face (mianzi). Victims who spoke out risked public shaming, disadvantages in the marriage market, and accusations of having invited assault~\cite{wang2017gender}. As W3-06 noted: \textit{``When facing safety threats such as sexual assault, cultural and societal factors often make women in China reluctant to report incidents to the police, largely because law enforcement lacks gender-sensitivity awareness when handling such cases. In addition, because many officers are male, they may struggle to approach these issues from a female perspective.''}

A few participants also pointed to the limited practical support provided by organizations such as the All-China Women’s Federation (Fulian), which are ostensibly positioned to advocate for women’s rights and protection but often face resource constraints and insufficient public attention. By contrast, some participants acknowledged that the female driver mode in Didi could significantly increase their sense of safety due to a reduced risk of harassment, enhanced trust in drivers, and the comfort of a gender-sensitive environment~\cite{cai2023ride}.

In addition, several participants invoked cultural practices and beliefs as informal mitigation measures. They noted that items such as temple talismans were placed at doorways or worn on the body to ``raise yang energy'' and guard against misfortune. However, as W3-04 explained, the traditional use of protective amulets provided psychological reassurance rather than practical protection. Family members of some participants viewed unmarried YWLA as ``heavy yin power'' and, therefore, as more vulnerable. W4-01 (from Northwest China) stated: \textit{``My grandmother brings me peace talismans and a peachwood sword from the temple and tells me to hang them on the door.''} In this case, these folk items functioned less as deterrents than as tangible expressions of familial care, reinforcing bonds of filial reciprocity, as W4-01 further explained: \textit{``[wearing these talismans] puts my parents at ease; they feel reassured knowing their daughter is protected.''}

Furthermore, regional norms shaped whether such practices were experienced as reassuring or uncanny. For instance, W4-02, from a different province in Northwest China, associated peachwood swords and embroidered shoes with funerary rites and confessed that she would \textit{``[...] be too scared to go home because these items have strong yin power''} if they appeared at her door. This contrast underscores that folk metaphysics (Xuanxue)\footnote{\url{https://en.wikipedia.org/wiki/Xuanxue}} is not universally embraced; its perceived efficacy is culturally situated and highly personal.


\subsubsection{Crowdsourced Peer Loops on Social Media} \label{mitigate_social_media}
Nearly all participants reported actively engaging with social platforms (e.g., RedNote, Douyin) as part of their evolving strategies to address or mitigate PSS concerns and cope with everyday vulnerabilities--for example, researching smart home device purchases or exchanging boundary-setting strategies in peer forums.

However, several participants also highlighted the emergence of discursive battles in the comment sections of PSS-related content, where gender debates often overshadowed the original safety topic. For instance, W5-01 and W5-02 discussed how videos of women being harassed, or surveillance footage showing suspicious behavior, frequently ignited threads that devolved into gender antagonism, with responses polarized between expressions of solidarity and overt misogyny. W5-02 further noted that attempts to engage in discussions were met with hostility or dismissiveness, underscoring how gendered PSS concerns were often politicized or invalidated in public discourse.

While social media networks offered empowerment and access to information, they simultaneously fostered heightened anxiety, disempowerment, and psychological strain~\cite{sampson2018affect}. Many participants noted that viewing PSS-related content triggered a flood of similarly themed posts controlled by algorithms, escalating from general safety tips to extreme, sensational narratives involving stalking, assault, and home invasions. W3-05 stated: \textit{``These videos often use unsettling background music, dramatized reenactments, and sensationalized titles, all of which contribute to a creepy atmosphere.''} Many participants described feeling trapped in echo chambers of fear, exacerbated by platform monetization structures that rewarded engagement driven by emotional arousal. These fear-based information loops often blurred the line between caution and alarmism.

Participants also expressed growing skepticism--and raised ethical concerns--about influencers who profited from sensational content. W3-05 and W5-07 acknowledged that, initially, exposure to such material increased their PSS awareness, but the realization that certain content creators might have fabricated or exaggerated events for visibility eroded trust. W5-07 explained: \textit{``I’ve seen extreme suggestions like buying a chainsaw or splashing boiling oil on intruders. That’s excessive and dangerous. If you misjudge the situation and overreact, it could lead to serious legal consequences. It’s important to stay rational and act within legal boundaries.''}

Furthermore, several participants (e.g., W4-01, W4-03) reported developing symptoms akin to hypervigilance, including intrusive thoughts, sleep disturbances, and distrust of public environments, ultimately leading them to selectively avoid or block certain types of content. To address these challenges, some participants demonstrated digital literacy and forms of resistance through \textit{``anti-algorithm''} tactics, intentionally searching for non-PSS content to reset their feed. They also engaged in content creation themselves, reposting safety knowledge with added commentary or corrections based on their own lived experiences.

\noindent\shadowbox{\parbox{\linewidth}{\textbf{Key Findings (RQ2):} Participants adopted four strategies to navigate the threats identified in \S\ref{threat_model}: 1) leveraging smart home devices (e.g., surveillance cameras and smart doorbells/locks) to bolster personal PSS; 2) performing masculinity and adjusting and managing privacy boundaries--such as placing men's items at home entrances and modifying social media privacy settings to reduce vulnerability; 3) utilizing traditional Chinese beliefs like talismans as psychological measures to feel safe and comfortable; and 4) relying on online community support forums for emotional resilience and practical advice.

However, participants critically pointed out significant drawbacks of each mitigation strategy, including persistent privacy vulnerabilities inherent in smart home ecosystems, constrained behavioral freedom due to patriarchal societal expectations, problematic perpetuation of purity norms specific to women, and anxiety exacerbation through sensationalized and algorithmically amplified social media content.}}%

\subsection{Desired Improvements and Implications} \label{implications_suggestions}

\textbf{Accountability for Data Handling.} Many participants emphasized that companies should be responsible for data governance. W3-05 suggested that regulators mandate real-time compliance and establish \textit{``robust complaint mechanisms.''} W5-07 proposed displaying dynamic privacy conformity scores on product listings and packaging, ensuring that any infringement triggered fines proportionate to the sensitivity of the compromised data--effectively creating a \textit{``Chinese GDPR-plus''} version with instant sanctions. Participants also highlighted challenges in addressing gender-specific PSS needs without inadvertently creating new risks. W5-05 noted that adding features specifically for women could backfire by exposing sensitive information. W5-06 questioned the assumption that women required special treatment, cautioning that marketing strategies emphasizing a product’s usefulness for women might inadvertently reinforce the notion of female vulnerability: \textit{``Women should not have to label men; replacing gender tags with a generic emergency trigger usable by any resident would make the tool unbiased and inclusive.''}

\textbf{Privacy Policy Restructuring.} Participants traced anxiety to complex privacy policies, describing them as \textit{``inscrutable black boxes''} (W2-03). They recommended layered, modular policies rather than binary consent designs and called for mandatory executive summaries containing only essential data practices (e.g., data types collected, retention duration, third-party data sharing). As W3-06 stated, summaries should be \textit{``limited to 200 characters or fewer and be easily understood at a 30-second glance.''} Participants emphasized the need for closer collaboration between compliance officers, UX designers, and engineers (W4-04). W1-02 further recommended contextual explanations and visual data flow diagrams instead of dense legal text. Additionally, W1-01 and W5-07 advocated for temporal data visualizations that clearly showed policy changes since the last acceptance--alongside risk-impact summaries of these modifications. W6-02 highlighted integrating readability scoring (e.g.,~\cite{pang2006chinese}) into policy development to improve comprehension and bridge the gap between legal obligations and user expectations.

\textbf{Branding PSS and Making Data Flows Visible.} Participants encouraged companies to explicitly brand their PSS engineering investments and embed safeguards directly in smart home devices, such as designing clear data deletion pathways to verify footage removal from clouds (W3-04). Such transparency would allow consumers to choose products based on verifiable security rather than aesthetics alone. W5-03 noted, \textit{``If one of two competing products prioritizes privacy, users will choose it, pushing the entire industry forward.''} Similarly, some participants proposed progressive disclosures through real-time data dashboards to enhance transparency and demonstrate measurable trust benefits. W4-04 described such dashboards as particularly valuable for smart locks: \textit{``A real-time activity log showing historical access, data collection details, retention periods, and personal preferences would improve the security perception of women living alone.''}

\textbf{Responsibility of the Government and Legal Reform.} Participants emphasized that the government bore primary responsibility, seen as greater than that of individuals or enterprises, highlighting its dual role as protector and regulator. The low cost of violations was identified as a critical barrier to effective deterrence. W3-02 stressed, \textit{``If laws become stricter and penalties harsher, incidents are likely to decrease significantly.''} Many participants called for substantially increased penalties for violations, particularly in cases involving gender-based violence, harassment, or digital exploitation. W3-04 and W6-02 noted that the current punitive model was insufficiently responsive to the severity of harms experienced by women, particularly emphasizing the need for stronger penalties and judicial consequences under the Public Security Management Punishment Law.\footnote{Article 49, Items 6–7 specify that individuals who repeatedly send pornographic, insulting, or threatening content, harass others, or invade personal privacy through actions like peeping, secret filming, or eavesdropping may be detained for up to 5 days or fined up to 1,000 RMB. In more serious cases, detention may extend to 10 days with the same fine~\cite{public_security_law}.} W1-01 further advocated for comprehensive, enforceable legal protections designed to safeguard YWLA's PSS rights proactively rather than reactively.

\textbf{Implementation, Monitoring, and Community Initiatives.} Beyond creating and reinforcing laws, participants highlighted the importance of effective implementation and monitoring--for example, building community-level management structures and deploying encrypted data systems for personal information, as W6-02 proposed. W4-03 shared positive experiences with localized safety initiatives, including prominently displaying police emergency contacts in public spaces and increasing female police visibility to enhance perceived safety. W3-04 further recommended leveraging technological advancements to strengthen public security: \textit{``Public surveillance systems should integrate facial recognition with criminal record databases to alert authorities when known offenders enter sensitive areas, proactively reducing potential threats.''}

Participants also emphasized the need for comprehensive, continuous education on privacy, security, and gender equality from an early age to address underlying societal issues contributing to security threats. Interestingly, a few participants cautioned against categorizing YWLA as inherently vulnerable, as this could inadvertently reinforce negative stereotypes. W6-06 emphasized, \textit{``I seek equal and universal protections, not special treatment.''}

\noindent\shadowbox{\parbox{\linewidth}{\textbf{Key Findings (RQ3):} Participants outlined comprehensive recommendations for improvements across three dimensions: legal and regulatory reforms, privacy governance, and technological enhancements: 1) emphasizing the state's primary responsibility for establishing robust legal frameworks, including substantially increased penalties for violations, particularly for gender-based digital and physical abuses; 2) institutionalizing community management systems, encrypted data practices for personal information, and regular safety checks, such as promoting gender equality from an early age, which were strongly endorsed to reduce gender biases and violence; 3) advocating clear, concise, and user-friendly privacy-policy presentations, incorporating executive summaries, visual aids, and contextual explanations.}}

\subsection{Iterative Development of PSS Guidebook} \label{pss_guidebook}
We developed an initial version of the PSS guidebook based directly on findings from our analysis of the PTM workshop transcripts. Drafted by the first author, this version was explicitly structured to address the key threats and mitigation strategies identified during the workshops. While many recommendations were drawn from participants' lived experiences, the content was curated and contextualized by the research team to ensure practical relevance and coherence. The guidebook was designed to support personal safety, digital privacy, and mutual aid, specifically tailored to the lived realities of Chinese YWLA. Drawing on recurring themes and participant input, it comprises seven clearly defined sections, each addressing a particular area of concern or strategy frequently discussed.

\begin{itemize}
\item 1) Physical threat prevention includes practical advice from participants, such as installing smart doorbells and cameras, and simulating the presence of male inhabitants (e.g., placing men’s shoes at entrances or using recordings of male voices).
\item 2) Digital security measures address harassment and scams (e.g., deepfake scams targeting family members), emphasizing data protection, family verification strategies, and social media privacy settings.
\item 3) Hidden camera detection provides guidance for identifying covert surveillance in shared spaces—a concern raised by participants—alongside recommendations for managing data from smart home devices.
\item 4) Building mental resilience, inspired by discussions on community support, encourages forming online and offline safety networks and recommends practices such as emergency hand signals and mutual aid within residential communities.
\item 5) Social media management tackles algorithm-driven anxiety loops, offering practical advice on managing exposure and identifying misleading or sensationalized PSS-related content.
\item 6) Reporting (including legal rights) highlights relevant legal frameworks, reflecting participants’ calls for clearer guidance on evidence preservation and formal avenues for recourse.
\item 7) Inspired by the ancient Chinese classic The 36 Stratagems,\footnote{The 36 Stratagems is a traditional Chinese text outlining tactical strategies historically applied in warfare, politics, and daily life: \url{https://en.wikipedia.org/wiki/Thirty-Six_Stratagems}.} we re-imagined its spirit of clever deception, boundary management, and strategic retreat into 36 detailed strategies, each paired with real-life scenarios and corresponding mitigation practices.
\end{itemize}


To evaluate the guidebook, we distributed it to the same 33 participants who took part in the PTM workshops and asked them to complete a follow-up survey with open-ended questions. This survey collected qualitative feedback, ensuring that participant insights were directly incorporated into the revisions. We received 31 valid responses, all of which confirmed the guidebook's clarity and usefulness, while several participants also offered constructive critiques and suggestions. For example, W4-03 recommended more detailed guidance on effective reporting channels (e.g., specific legal articles), and W2-01 called for the inclusion of local support networks or hotlines. Several participants (W1-02, W3-05, W4-02, W6-03) suggested adding video demonstrations related to hidden camera detection, and W5-07 requested advice on identifying misinformation on social media.%

Additionally, although a small number of participants found the guidebook lengthy, this view was not shared by the majority. Analyzing responses to the fourth open-ended question, we found that most participants felt safe and reassured after reading the guidebook, noting that it was detailed, covered many real-life scenarios, and was made more engaging through its connection to traditional Chinese stories (the 36 Stratagems).

Based on this feedback, we revised and expanded the guidebook. Revisions included adding concrete legal articles with brief explanations in the legal section; incorporating official video resources related to hidden camera detection (e.g., from government or public security accounts); and expanding the social media section with real-world case studies to illustrate misinformation identification. The final guidebook was published on a publicly accessible website, featuring AI-generated images and interactive, user-friendly interfaces. It was shared through social media platforms (e.g., RedNote) to ensure broader reach and practical utility beyond the original participant group, thereby fulfilling our research aim of providing tangible support for Chinese YWLA.



\section{A Human-Centered Threat and Mitigation Framework for Chinese YWLA} \label{threat_model_framework}
Drawing on our empirical insights in \S\ref{threat_model} and \S\ref{mitigation_practices}, we conceptualize our findings as a human-centered threat model for Chinese YWLA, inspired by Usman and Zappala's framework on human-centered threat modeling~\cite{usman2025sok}, which emphasizes four components—\emph{context}, \emph{threats}, \emph{protective strategies}, and \emph{reflection}—to move beyond systems-centric models~\cite{deng2011privacy,wuyts2020linddun}. We instantiate this structure for YWLA living in smart home environments under Chinese patriarchy. A defining contextual feature of our population is a ``guardianship gap'': participants occupied apartments and digitally mediated neighborhoods without partners, roommates, or family members who might otherwise help protect against unwanted contact (\S\ref{threat_model}). Rather than treating these threat factors in isolation, we show how they interact and reinforce one another: platform and recommendation system leaks of addresses facilitated offline violence; pervasive state and corporate surveillance normalized intrusive monitoring; and gendered social norms influenced perceptions of who could be considered a credible victim or offender. This cycle echoes Thomas et al.'s findings on how different forms of hate and harassment cross-amplify online and offline harms~\cite{thomas2021sok}, but situates it specifically within the lived experiences of YWLA in China.

Regarding mitigation strategies (\S\ref{mitigation_practices}), we also extend Warford et al.'s work by treating mitigations not as unambiguously beneficial countermeasures but as double-edged configurations that can simultaneously reduce some harms while amplifying others for YWLA~\cite{warford2022sok}. For example, smart cameras and locks acted as ``digital guardians,'' enabling participants to monitor doors and corridors, yet they also expanded the circle of actors (e.g., landlords) who could access intimate data. Performative masculinity or strategic self-presentation on social media could deter opportunistic actors while reinforcing heteronormative assumptions that a woman alone was inherently unsafe. Similarly, social media communities provided mutual aid and threat intelligence but were driven by algorithms that over-surface fear-inducing content, entrenching a pervasive sense of ambient threat.%

\subsection{Rethinking Privacy Ownership and Boundaries} \label{rethinking_CPM}

To further articulate how threats and mitigation strategies reshape control over PSS, we draw on CPM theory~\cite{petronio1991communication,petronio2020conceptualization}. CPM conceptualizes privacy as the ongoing management of informational and spatial boundaries across relationships, emphasizing three key ideas: \emph{privacy ownership} (the belief that individuals own and should control their personal information), \emph{privacy control} (the rules governing when, how, and to whom information is disclosed), and \emph{privacy turbulence} (moments when those rules are violated or destabilized, requiring boundary repair or renegotiation). Framing our human-centered threat and mitigation model through the lens of CPM allows us to extend existing threat modeling in two key ways. First, we show how the threats and responses of YWLA in China are shaped by shifting regimes of co-ownership and boundary negotiation, where smart homes, digital platforms, landlords, and the state (among other actors) all assert control over personal data. Second, we demonstrate that mitigation strategies employed by YWLA extend beyond reducing technical attack surfaces to redefining who counts as a co-owner of data, clarifying their responsibilities, and addressing boundary turbulence as it arises.

Our findings in \S\ref{threat_model} reveal that YWLA face a ``guardianship gap,'' in which traditional co-owners (e.g., family members or partners) are absent, while new, often uninvited co-owners emerge, such as landlords or male roommates with access to keys and shared spaces. Incidents like address leakage through delivery apps or hidden cameras in rentals exemplify repeated episodes of boundary turbulence, YWLA's privacy rules (e.g., protecting their location or intimate spaces) are frequently violated. Living alone intensifies these disruptions, as YWLA must shoulder the cognitive and emotional burden of boundary repair, including negotiating with others or avoiding confrontation with delivery workers~\cite{kang2023communication}. Furthermore, the mitigation strategies described in \S\ref{mitigation_practices} are not ad-hoc safety tips but systematic efforts to redraw privacy boundaries under the recurring turbulences across domestic, platform, and state levels~\cite{meng2024interpersonal}. Smart home devices allow YWLA to reconfigure household boundaries by turning locks, cameras, and logs into tools for screening visitors, monitoring corridors, and gathering evidence. Boundary management tactics, including sitting in the back seat of ride-hails, avoiding confrontation, adopting male names or avatars online, and masking femininity through clothing, reflect deliberate attempts to adjust boundary permeability in relation to different male actors.

These boundary dynamics resonate with broader accounts of surveillance capitalism and digital patriarchy. Commercial and state infrastructures do not merely ``collect data,'' they extract behavioral traces and access logs in ways that reinforce gendered control over women's mobility and self-presentation, positioning platforms, landlords, and grid managers as privileged interpreters of YWLA's threats~\cite{Zuboff2019}. In our human-centered threat model, smart home vendors, digital platforms, and authorities like grid-management systems simultaneously act as service providers and as safeguards capable of overriding YWLA's own privacy rules, illustrating how data-driven safety infrastructures may reproduce, rather than dismantle, patriarchal power~\cite{stardust2025s}.

\subsection{Positioning YWLA Threats Within Hate and Harassment Frameworks}

Although participants did not necessarily self-identify as at-risk, our study falls within the scope of at-risk user research because the threats we examined are closely intertwined with retaliation risks, cross-channel escalation, and limited avenues for recourse~\cite{warford2022sok,bellini2024sok}. Our findings also align with, and refine, existing hate and harassment taxonomies~\cite{thomas2021sok}, showing how harassment categories and harm pathways are amplified across online and offline settings. Participants rarely experienced incidents as isolated forms of ``online abuse'' or ``offline danger''; instead, they described connected trajectories in which platform-mediated data leakage, sexually harassing contact, and intimidation in physical spaces reinforced one another, potentially leading to downstream harms such as stalking or family-targeted fraud.

This framing also helped us distinguish which dynamics generalized across contexts and which were specific to our participants. Several findings echoed widely documented gendered safety-related issues and topics, including routine sexual harassment, fear of stalking, and boundary work through self-protective routines and peer support~\cite{liu1999enduring,xu2005prevalence,wang2023some,Zeng2020,sugiura2020victim,he2025living}. At the same time, our findings highlighted mechanisms particularly relevant to our participants, including: (1) a guardianship gap shaped by living away from kins while remaining relationally accountable to parents~\cite{gui2020leftover,ji2015between,gaetano2014leftover,yu2011varying,ge2014changes}; (2) a governance environment in which online safety discourse and feminist organizing communities were simultaneously enabled and constrained through heavy policing, boundary work, and misogynistic backlash~\cite{xu2011internet,peng2020feminist,online_space_han_2018}; and (3) the framing of feminism as a Western import, which further narrowed the perceived legitimacy of safety claims~\cite{huang2023anti,yang2023rural,mao2020feminist,yin2021intersectional,deng2024persuasion,wu2019made,peng2020feminist}. These context-specific dynamics do not negate cross-context patterns; rather, they illustrate how the same categories of harassment and coercion can become more or less feasible, reportable, and repairable depending on local norms and institutional arrangements.%

\section{Design and Policy Recommendations and Implications} \label{design_suggestions}

\subsection{Smart Homes as Infrastructures Centering YWLA PSS} \label{suggestion_to_smart_home}

Drawing on \S\ref{mitigate_smart_home}, we highlight the need for clear, intuitive data management settings embedded directly into smart home devices, such as activity logs and access histories that allow users to detect suspicious behavior. However, our data suggests that YWLA do not merely need more data; they need abstractions that align with their lived threat model. Accordingly, we propose scenario-based ``home alone'' safety modes that reconfigure devices around concrete risk situations (e.g., being home alone at night) rather than generic security levels. For example, a ``home alone at night'' mode could, by default, tighten door-lock authentication, temporarily extend doorstep video retention, and prioritize rapid-access panic workflows and short-term encrypted local storage. While participants requested real-time dashboards and deletion controls as prior studies have suggested~\cite{mare2020smart,marky2024decide,windl2023investigating,thakkar2022would}, they also described exhausting routines of repeatedly checking cameras and logs. To address this, we suggest dashboards that reduce cognitive and labor burdens by surfacing a small number of prioritized alerts and actions, focusing on events most relevant to the user's immediate safety concerns (e.g., unfamiliar access). Rather than listing every interaction, the dashboard interface could highlight anomalies relative to the user's routine and offer one-click actions, such as switching to ``home alone'' mode.

Second, some participants relied on physical props, such as men's shoes, to deter threats (\S\ref{mitigate_relationship}). We suggest that smart home product teams explore ways to simulate occupancy that go beyond simple light timers. For instance, devices could automatically play recordings of male voices or cast shadows of movement onto curtains when a doorbell rings or a stranger lingers. This shifts the burden of performing safety from the user to the system, allowing YWLA to leverage the deterrent effect of a ``male presence'' without needing to stage it manually. Additionally, smart home teams can integrate interactive tutorials and scenario-based privacy training during device setup (e.g., using visual aids or animations during in-app registration and activation), empowering less tech-savvy YWLA to make informed decisions about device configurations, data-sharing preferences, and remote access settings.

Finally, participants called for companies to make investments in PSS legible as a basis for product choice (see \S\ref{mitigate_smart_home} and \S\ref{implications_suggestions}). We emphasize the need to transform existing documentation into concise, modular, and visually accessible formats. For example, summaries capped at 200 characters, combined with readability scores and icon-based explanations, should replace dense, legalistic texts that discourage user engagement. Building on prior work on privacy labels~\cite{koch2022keeping,kollnig2022goodbye,li2022understanding,zhang2022usable,kelley2009nutrition}, we propose YWLA-centered threat model audits and trustmarks that explicitly certify how a device performs under a YWLA-risk profile (e.g., resistance to address leakage, support for bystander privacy). Rather than generic security-certified seals, such labels would summarize how the design fares against the concrete actors, attacks, and harms in our framework.

\subsection{Social Media Platforms as PSS Ecologies} \label{suggestion_social_media}

Our findings (see \S\ref{mitigate_social_media}) indicate that viral reports of violence against Chinese YWLA create echo chambers that amplify perceptions of vulnerability and reinforce aggressive protection narratives~\cite{cinelli2021echo,he2025living}. Posts promoting extreme safety measures, expressing anger, or sharing sensational screenshots often receive high engagement, fostering solidarity among women while normalizing surveillance as a necessary solution. This links empowerment to the purchase of surveillance devices and reinforces the notion that autonomy can be achieved through these tools~\cite{sadowski2020too,thornham2010just,wu2019made,yang2023post,yang2024girls,he2025living}. Such discourse can also fuel antagonism toward men and stigmatize feminism, ultimately undermining efforts to achieve genuine gender equality~\cite{yang2023rural,deng2024persuasion}.

To address these dynamics, we suggest that social media platforms improve algorithmic recommendation mechanisms, enabling users to understand and adjust how their viewing histories shape recommendations (e.g., through ``anti-algorithm'' tactics; see \S\ref{mitigate_social_media}). Our recommendations go beyond abstract calls by specifying how algorithms could respond to a human-centered threat model. We further propose a dual-layer governance model: one layer ensures regulatory compliance,\footnote{~\href{https://www.gov.cn/zhengce/zhengceku/2022-01/04/content_5666429.htm}{Provisions on Administration of Algorithmic Recommendation in Internet Information Services}} while the other prioritizes user safety through robust content moderation. This safety layer should down-rank sensational or misleading personal-security advice and limit reposting chains that perpetuate fear. Such a framework respects mandated content takedowns while addressing algorithmically driven anxiety loops and reducing the stigma surrounding Chinese feminist discourse. For example, recommendation systems could monitor sentiment in PSS-related posts and, if overwhelmingly negative, offer users tools—such as a slider—to reduce the prominence of highly emotive content. Rather than simply hiding violent content, this approach acknowledges that YWLA often seek PSS information but need safeguards to prevent algorithmic amplification from turning legitimate vigilance into anxiety.

Drawing on \S\ref{scam_threats}, we suggest embedding family-centered verification rituals into social platforms. Platforms could allow users to co-create verification protocols with older relatives (e.g., pre-agreed questions or codes) that are automatically triggered when a high-risk request is detected, such as an urgent money transfer above a predefined threshold. This approach translates culturally situated filial obligations into concrete interface support and operationalizes the protective strategies dimension of human-centered threat modeling at the family-network level. In addition, participants reported feeling overwhelmed by contradictory PSS advice (see \S\ref{mitigate_social_media}). Rather than merely embedding generic suggestion modules, we propose prioritization overlays for PSS-related advice and content that help YWLA evaluate such information along multiple axes. For example, when a user encounters or shares a PSS-related post, the platform could display a standardized pop-up or panel indicating the evidentiary basis of the advice, endorsements by vetted organizations, and potential trade-offs (e.g., the risk of escalating conflict with neighbors). These interface elements could be co-designed with feminist groups and at-risk communities to reflect context-specific norms, shifting platforms from passive amplifiers of sensational advice to active curators of situated, high-quality safety practices.

\subsection{Suggestions for Policymakers} \label{policy_suggestions}
\subsubsection{Social Perspective}

Building on \S\ref{mitigate_cultural}, we extend He et al.’s work~\cite{he2025living} on culturally specific PSS mitigation practices by revealing how traditional beliefs intersect with contemporary PSS strategies in China. These findings advance prior HCI research by foregrounding the importance of socio-cultural context from a non-Western perspective and by underscoring the need for culturally responsive approaches to the design of PSS tools and educational materials~\cite{he2025exploring,bailey2024utilizing,albayaydh2024co}. Drawing on prior work~\cite{warford2022sok,usman2025sok}, we argue that such cultural practices should not be treated as peripheral add-ons, but rather as starting points for institutionalized, participatory risk assessment. Concretely, authorities could incorporate recurring, human-centered threat modeling sessions with YWLA into grid-management and residents’ committee activities, treating YWLA as expert witnesses whose lived experiences inform local “risk registers” for both digital (e.g., public CCTV) and physical (e.g., security patrols) infrastructures. In parallel, we advocate for participatory mechanisms that enable Chinese YWLA to communicate concerns directly to community representatives, including training grid workers and residents’ committee staff in gender-sensitive engagement.

To raise public awareness of digital scams and deepfakes, government agencies should collaborate with technology companies and local organizations to launch public-awareness campaigns that also address the stigmatization faced by sexual harassment victims, fostering supportive and non-judgmental community responses. We further recommend increased resourcing for organizations such as the All-China Women’s Federation, along with greater representation of female staff to reduce gender-based distrust. Beyond general support, these organizations could be formally mandated as “PSS intermediaries” to (1) operate confidential reporting channels for YWLA who fear retaliation from landlords, device vendors, or digital platforms, and (2) broker access to legal, psychological, and technical support, such as assistance with evidence collection or deepfake verification.
\subsubsection{Legal Perspective}

We recommend that policymakers strengthen the Public Security Management Punishment Law by substantially increasing penalties for digital harassment, stalking, deepfakes, and gender-based violence (see \S\ref{implications_suggestions}). Legal reforms should also introduce clearer definitions and explicit protections for victims of digital violence, shifting the legal emphasis from victim-blaming toward victim support~\cite{henry2016sexual}. In parallel, we call for the integration of a more robust data protection framework within the existing PIPL, with clearly defined and immediately enforceable sanction mechanisms. Such a framework should mandate real-time compliance monitoring, enhance transparency, and establish strong corporate accountability to prevent data misuse and better safeguard user privacy.

To address the “burden of proof” problem described by participants in deepfake and harassment cases (\S\ref{threat_model}), legal reforms could further introduce presumptive protections.\footnote{\href{https://www.gov.cn/zhengce/zhengceku/202307/content_6891752.htm}{Interim Measures for the Management of Generative Artificial Intelligence Services 2023 (2023 AIGC Interim Measures)}} For example, sexually explicit synthetic media could be treated as non-consensual by default unless robust evidence of consent is provided. This approach would align Chinese practice with emerging international debates on digital sexual violence while grounding legal doctrine in locally salient harms, such as filial blackmail and reputational damage to families.

\subsubsection{Educational Perspective}

We advocate for integrating comprehensive digital literacy programs into school curricula, with our guidebook serving as a pedagogical resource to support teaching and learning. Overseen by educational authorities, these programs should address PSS concerns while promoting gender equality, covering topics such as deepfake identification, responses to online harassment, and the responsible use of smart home technologies. Together, these components can equip younger generations with the skills needed to navigate digital environments safely and confidently. In parallel, policymakers should promote gender equality through sustained, community-level educational initiatives that challenge entrenched societal biases and encourage respectful interaction from an early age~\cite{wang2024attitude,ferrer2021bias}. In line with~\cite{redmiles2020comprehensive}, such curricula should not merely enumerate long lists of safety tips, but should instead prioritize advice. Building on our co-developed PSS guidebook as an initial blueprint, educational authorities can adapt and iteratively refine its scenario-based modules—rather than designing materials in isolation—to better reflect institutional constraints and local practices.

In addition, drawing on insights from \S\ref{mitigate_cultural}, we emphasize the importance of professional training for law enforcement and judicial personnel, particularly when handling cases of digital and physical harassment. Given the cultural emphasis on female purity in Chinese society, such training can foster a deeper understanding of victims’ experiences, improve institutional responses, encourage reporting, and build trust in the judicial process~\cite{lorenz2024sexual}. To avoid overly generic training, we recommend using anonymized cases derived from YWLA’s threat modeling outputs (e.g., escalation from online harassment to physical stalking via address leakage) and evaluating officers on their ability to minimize secondary victimization (e.g., avoiding moral judgment), rather than relying solely on case-closure metrics.

\section{Conclusion}
We conducted six PTM workshops with 33 Chinese YWLA, revealing how three primary threat types are interconnected, as well as how cultural practices, social media, and smart home technologies shape both the perception and mitigation of PSS risks. We surfaced a human-centered threat model in which digitally facilitated physical violence, digital harassment and scams (e.g., deepfakes), and multi-scalar surveillance are mutually reinforcing, shaped by patriarchal institutions. We showed how four mitigation strategies--smart home device use, boundary and relationship management, cultural practices, and social media engagement--are experienced as double-edged, often redistributing rather than addressing risks. We also integrated the CPM framework into our human-centered threat model, clarifying how YWLA share privacy co-ownership and boundaries. This revealed how boundary turbulence arises and how these breaches are repaired in daily life. Furthermore, our work contributes to feminist HCI and usable security and privacy literature by foregrounding culturally specific threat responses and offering actionable recommendations for design, governance, and participant-centered security and privacy practices. Ultimately, we translated our findings into a practical guidebook to help Chinese YWLA navigate and address or mitigate their PSS concerns in daily life.

\begin{acks}
We thank all participants for their time and valuable contributions. We also appreciate the constructive feedback from the anonymous ACM CHI reviewers, which helped enhance the clarity and presentation of our work. ChatGPT was used to assist with proofreading and grammar checking. This research was partially funded by the Spanish Government under grant PID2023-151536OB-I00 and by the Generalitat Valenciana through the PROMETEO project CIPROM/2023/23.
\end{acks}
\bibliographystyle{ACM-Reference-Format}
\bibliography{reference}

\end{document}